\documentclass[11pt]{article}
\usepackage{amssymb,amsmath,amsfonts}
\usepackage{graphicx}
\usepackage{color,epsfig}
\usepackage{graphics}
\usepackage{eepic,epsfig}

\begin{document}

\title{Induced current by a magnetic flux in $(1+2)-$dimensional conical spacetime in a Ho{\v{r}}ava-Lifshitz Lorentz-violating scenario}

\author{E. R. Bezerra de Mello$^1$\thanks{emello@fisica.ufpb.br} \ and
	H. F. Santana Mota$^1$\thanks{hmota@fisica.ufpb.br} 
	\\
	\textit{$^{1}$Departamento de F\'{\i}sica, Universidade Federal da Para\'{\i}ba,}\\
	\textit{58.059-970, Caixa Postal 5.008, Jo\~{a}o Pessoa, PB, Brazil}}

\maketitle
\begin{abstract}
We investigate the vacuum expectation value of the bosonic current induced by a magnetic flux in a $(2+1)$-dimensional conical spacetime in the presence of a circular boundary within a Ho\v{r}ava-Lifshitz Lorentz-violating scenario. The circular boundary is assumed to be concentric with the magnetic flux, and the massive scalar field is taken to satisfy Robin boundary conditions on it. To carry out the analysis, we construct the positive-frequency Wightman functions in both regions, namely, inside and outside the circular boundary. Using these functions, we derive analytical expressions for the corresponding vacuum currents. We show that the Wightman functions, as well as the induced currents, can be naturally decomposed into boundary-free and boundary-induced contributions. For the boundary-induced current, several asymptotic behaviors are examined in relevant limiting cases. In addition, numerical results are presented in order to provide a clearer understanding of the dependence of the induced currents on the parameters of the model.
\end{abstract}

\bigskip


\section{Introduction}
\label{Int}
Topological defects are produced during phase transitions involving spontaneous symmetry breaking and play an important role in many areas of physics. They are present in different condensed matter systems, including superfluids, superconductors, and liquid crystals. In addition, symmetry breaking has also several cosmological consequences and, within the framework of grand unified theories, several types of these defects are predicted to have formed in the early universe \cite{Kibble,V-S}. They provide an important link between particle physics and cosmology. Among these defects, cosmic strings are of particular interest. They are candidates for producing interesting physical effects such as gamma-ray bursts \cite{Berezinski}, gravitational waves \cite{Damour}, and high-energy cosmic rays \cite{Bhattacharjee}. More recently, cosmic strings have attracted renewed interest, partly because a variant of their formation mechanism has been proposed within the framework of brane inflation \cite{Sarangi}-\cite{Dvali}.

In the simplest theoretical model describing an infinite straight cosmic string, the corresponding spacetime is locally flat except at its core, where the Riemann curvature tensor has a Dirac delta-function singularity. In this model, the two-geometry orthogonal to the string exhibits a planar angle deficit. From the point of view of quantum field theory, the corresponding nontrivial topology induces nonzero vacuum expectation values for several physical observables.
Explicit calculations associated with a single idealized cosmic string have been carried out for different fields \cite{Hell86}-\cite{Beze10}. In the presence of a very thin magnetic flux running along the string core, another important observable in this system is the induced vacuum current density, $\langle j^\mu \rangle$, associated with charged fields. This phenomenon has been investigated for massless \cite{LS}, massive scalar \cite{SNDV}, and fermionic \cite{ERBM} fields. In these works, the authors showed that a nonvanishing vacuum current density along the azimuthal direction arises when the ratio of the magnetic flux to the quantum flux has a nonzero fractional part. Moreover, the induced bosonic current in a higher-dimensional compactified cosmic string spacetime was calculated in \cite{Braganca2015}.

The presence of boundaries also produces modifications in the vacuum expectation values of physical observables. This phenomenon is a consequence of changes in the physical properties of the vacuum state. This is the well-known Casimir effect. The study of Casimir effects associated with an idealized cosmic string spacetime has been carried out for scalar, fermionic, and vector fields in \cite{Mello,Mello1,Aram1}, assuming specific boundary conditions on a cylindrical surface. In this context, the analysis of the induced bosonic current in a $(1+D)$-dimensional cosmic string spacetime, in the presence of a cylindrical boundary coaxial with the string, was developed in \cite{Mello:2025}. There, it was shown that, in addition to the standard induced azimuthal current, an extra contribution arises due to the presence of the boundary.

Lorentz invariance, which plays an important role in Quantum Field Theory, was called into question in a work by V. A. Kostelecký and S. Samuel \cite{Kostelecky_88}, who proposed a mechanism within string theory that allows for the violation of Lorentz symmetry at the Planck energy scale. According to this mechanism, Lorentz violation is induced by nonvanishing vacuum expectation values of certain vector and tensor fields, which select preferred directions in spacetime. In the context of quantum gravity, Ho\v{r}ava-Lifshitz (HL) proposed a theory \cite{Horava_09} with the aim of improving the ultraviolet behavior of quantum corrections by introducing different scaling properties for space and time coordinates. The HL approach explicitly breaks the symmetry between space and time and can be applied not only to gravity, but also to other field-theoretical models, including scalar, spinor, and gauge theories.
  
Lorentz-symmetry violation has become a subject of great experimental interest. For instance, highly precise measurements of the Casimir pressure have become an important tool in the investigation of Lorentz-violating quantum field theoretical models. In this context, Casimir systems provide a useful theoretical laboratory for indirectly exploring possible manifestations of Lorentz-symmetry violation in quantum vacuum phenomena. The first analyses of the Casimir effect in Lorentz-violating (LV) theories were carried out in \cite{FrankTuran,Escobar}, considering different Lorentz-breaking extensions of QED. In addition, the Casimir effect associated with massless scalar and fermionic quantum fields confined between two large parallel plates, within the Ho\v{r}ava-Lifshitz formalism, was investigated in Refs.~\cite{Ulion:2015kjx} and \cite{Deivid}, respectively. More recently, the Casimir effect associated with a massive fermionic field was studied in \cite{deMello:2024}.

In the present paper, we continue the investigation of the effects caused by Lorentz violation within the HL scenario. Specifically, we analyze the induced bosonic current in a $(1+2)$-dimensional conical spacetime in the presence of a magnetic flux located at the apex of the cone and a concentric circular boundary. Our main objective is to calculate the vacuum currents in the regions inside and outside the circular boundary. Having obtained the corresponding analytical expressions, we then investigate their main physical properties and discuss their most important features.

This paper is organized as follows. In Section \ref{sec2}, we present the background geometry associated with the conical planar system and the explicit expression for the three-vector potential associated with the magnetic flux. Considering the modified Klein-Gordon equation and the Robin boundary condition imposed on the charged field at the circular boundary, we calculate the complete set of normalized wave functions. By using the mode-summation formula, we obtain the positive-frequency Wightman functions for both regions of the spacetime. As we will see, the corresponding Wightman functions are expressed as the sum of a boundary-free contribution and a boundary-induced one.
In Section \ref{sec3}, we investigate the vacuum bosonic currents in both regions. There, we show that only azimuthal currents are induced. We also present some asymptotic behaviors of the boundary-induced currents.
In Section \ref{conc}, we summarize the main results obtained in this work. Throughout the paper, we use units such that $\hbar = G = c = 1$.

\section{Wightman function}
\label{sec2}
In this section, we analyze the quantum behavior of a massive charged bosonic field in a $(1+2)$-dimensional conical spacetime, considering the presence of a magnetic flux located at the apex of the cone and a circular boundary concentric with the flux. We assume that the field obeys Robin boundary conditions on the circular boundary. The main objective of this section is to calculate the positive-frequency Wightman functions for both regions of the space.

\subsection{Bulk geometry and bosonic modes}

In this subsection, we introduce the $(1+2)$-dimensional background spacetime with a conical geometry that will be considered throughout this paper, described by the line element
\begin{equation}
	ds^{2}=g_{\mu\nu}dx^{\mu}dx^{\nu}=dt^{2}-dr^{2}-r^{2}d\varphi
	^{2} \ ,  \label{ds21}
\end{equation}
with cylindrical coordinates $r\geqslant 0$, $0\leqslant \varphi\leqslant{2\pi}/{q}$, and $-\infty <t<+\infty$. The conical two-geometry is encoded by the parameter $q\geq 1$.\footnote{In the three-dimensional case, $q$ is related to the linear mass density of the cosmic string, $\mu_0$, by $q^{-1}=1-4\mu_0$.}

We also introduce the theoretical model in which the Klein-Gordon equation is modified within the framework of HL theory. Thus, the modified Klein-Gordon equation reads
\begin{eqnarray}
	\label{Mod_KG}
\left[\partial_0^2+l^{2(\xi-1)}\left(g^{ij}D_iD_j\right)^\xi+m^2\right]\phi(x)=0 \ ,
\end{eqnarray}
where we consider a massive charged scalar field coupled to a magnetic flux located at $r=0$. The parameter $\xi$ is the critical exponent, and for $\xi>1$ Lorentz symmetry is violated. Moreover, the parameter $l$ has dimensions of length. The magnetic flux is described by the three-vector potential
\begin{eqnarray}
	A_\mu=-\frac{q\Phi}{2\pi}\delta_\mu^2 \ ,
\end{eqnarray}
where $\Phi$ denotes the magnetic flux. Consequently, the covariant derivative is given by $D_{\mu}=\nabla_{\mu}+ieA_{\mu}$.

In addition, we impose that the solutions of the modified Klein-Gordon equation satisfy the Robin boundary condition on a circle of radius $a$:
\begin{equation}
	\left( A+B{\partial_r }\right) \phi =0,\quad r=a.
	\label{Dirbc}
\end{equation}
Hence, all results presented below depend on the ratio between the coefficients appearing in this boundary condition. However, in order to keep the transition to the Dirichlet ($B=0$) and Neumann ($A=0$) cases transparent, we employ the more general condition (\ref{Dirbc}).

Due to the cylindrical symmetry of the system, we assume that the general solution of Eq.~\eqref{Mod_KG} has the form
\begin{eqnarray}
\label{Sol_1}
\Phi_\sigma(x)=C_\sigma e^{-i(Et-qn\varphi)}R(r) \ , \ {\rm with} \ n=0, \pm1, \pm2, \ldots
\end{eqnarray}
where $\sigma$ denotes the set of quantum numbers and $C_\sigma$ is the normalization constant.

For the case $\xi=1$, the radial function is given by
\begin{eqnarray}
\label{Radial_sol}
R(r)=C_1J_{q|k_n|}(\lambda r)+C_2Y_{q|k_n|}(\lambda r) \ ,
\end{eqnarray}
where $\lambda$ corresponds to the radial momentum. In Eq.~\eqref{Radial_sol}, $J_\mu(z)$ and $Y_\mu(z)$ denote the Bessel and Neumann functions of order $\mu$, respectively \cite{Abra}. The order of these functions is
\begin{eqnarray}
k_n=n+\alpha \ ,
\end{eqnarray}
with $\alpha$ defined as the ratio of the magnetic flux $\Phi$ to the quantum flux $\Phi_0=2\pi/e$:
\begin{eqnarray}
\label{alpha_para}
\alpha=\frac{eA_2}{q}=-\frac{\Phi}{\Phi_0} \ .
\end{eqnarray}

In our analysis, we assume that $\xi$ is an integer. Since in Eq.~\eqref{Mod_KG} the spatial differential operator appears raised to a power greater than unity, the radial solution remains the same as in Eq.~\eqref{Radial_sol}. However, the corresponding dispersion relation becomes
\begin{eqnarray}
E_\sigma^2=l^{2(\xi-1)}\lambda^{2\xi}+m^2 \ .
\end{eqnarray}
In this case, $\sigma={n,\lambda}$.

The properties of the vacuum state can be described in terms of the positive-frequency Wightman function. This function is defined as the vacuum expectation value (VEV) of the product of the field operators,
\begin{eqnarray}
W(x,x')=\langle 0|\hat\phi(x)\hat\phi^\dagger(x')|0 \rangle \ ,
\label{Wigh_func_1}
\end{eqnarray}
Expressing the field operator in terms of creation and annihilation operators, Eq.~\eqref{Wigh_func_1} can be written in the mode-sum form
\begin{equation}
W(x,x')=\sum_{\sigma}\phi_{\sigma}(x)\phi_{\sigma}^{*}(x') \ .
\label{W.function}
\end{equation}

\subsection{Wightman functions}

In this section, we present the formal expressions for the Wightman functions in both regions of the spacetime, namely, inside and outside the circular boundary.

\subsubsection{Interior region}
\label{int_reg}

Assuming that the wave function satisfies the Dirichlet boundary condition at the origin, $r=0$, the solution given in Eq.~\eqref{solu1} takes the form
\begin{eqnarray}
\label{solu1}
\phi_\sigma(x)=C_\sigma J_{q|n+\alpha|}(\lambda r)e^{-i(E t-qn\varphi)}  \ ,
\end{eqnarray}

According to the Robin boundary condition, Eq.~\eqref{Dirbc}, the eigenvalues of the quantum number $\lambda$ are given by the solutions of the transcendental equation below:
\begin{eqnarray}
	{\bar{J}}_{q|n+\alpha|}(\lambda a)=0 \  ,
	\label{Cond.1}
\end{eqnarray}
where, for any function $f(z)$, we adopt the following notation,
\begin{equation}
	\bar{f}(z)=Af(z)+\left( B/a\right) zf^{\prime }(z)  \  .  \label{fbar}
\end{equation}

For a fixed value of $\alpha$, and a given $n$, the possible values of $\lambda$ in \eqref{Cond.1} are determined by the relation
\begin{equation}
	\lambda =\gamma _{\nu_n,i}/a,\quad i=1,2,\cdots ,  \label{ganval}
\end{equation}%
where, for convenience, we adopt the notation $\nu_n=q|n+\alpha|$. The positive roots of \eqref{Cond.1}, $\gamma _{\nu_n,i}$, are arranged in ascending order: $\gamma_{\nu_n,i}<\gamma_{\nu_n,i+1}$, for $i=1,2,\ldots$.

The normalization coefficient $C_\sigma$ is obtained from the condition
\begin{eqnarray}
	\label{Norm1}
	\int_0^a d^2x\sqrt{|g|}\phi^*_{\sigma'}(x)\phi_\sigma(x)=\frac{\delta_{i,i'}\delta_{n,n'}}{2E_\sigma} \  .
\end{eqnarray}
After some algebraic manipulations involving Bessel functions, we obtain
\begin{equation}
	|C_{\sigma}|^2=\frac{q}{2\pi E_\sigma}\frac{\gamma_{\nu_n,i}}{a^2}T_{\nu_n,i}(\gamma_{\nu_n,i})\ ,
	\label{Const_1}
\end{equation}
where the function $T_{\nu_n}(z)$ is given by
\begin{eqnarray}
	T_{\nu_n}(z)=z[(z^2-{\nu_n}^2)J_{\nu_n}^2(z)+z^2(J_{\nu_n}^{\prime}(z))^2]^{-1} \ .
\end{eqnarray}

Now we are in a position to calculate the Wightman function.
Substituting the explicit expressions for the normalized bosonic wave functions, we obtain:
\begin{eqnarray}
	\label{W_inside}
W(x,x')&=&\frac{q}{2\pi a^2l^{\xi-1}}\sum_{n=-\infty}^{+\infty}  \sum_i\frac{\gamma_{\nu_n,i}T_{\nu_n,i}(\gamma_{\nu_n,i})}{\sqrt{(\gamma_{\nu_n,i}/a)^{2\xi}+\mu^{2\xi}}}\nonumber\\
&\times&J_{\nu_n}(\gamma_{\nu_n,i} r/a)J_{\nu_n}(\gamma_{\nu_n,i} r'/a) e^{-i[E_\sigma(t-t')-qn(\varphi-\varphi')]}  \ ,
\end{eqnarray}
where we have explicitly substituted the expression for the energy in the denominator and replaced the mass of the field by $m=\mu^\xi l^{\xi-1}$.

In order to perform the summation over the quantum number $i$, we will use a variant of the generalized Abel-Plana summation formula \cite{Saha87}
\begin{eqnarray}
	\sum_{j=1}^{\infty }T_{\nu_n}(\lambda _{\nu_n,j})f(\lambda _{\nu_n,j}) &=&\frac{1}{2}\int_{0}^{\infty}dz\,f(z)-\frac{1}{2\pi }\int_{0}^{\infty }dz\,\frac{\bar{K}_{\nu_n}(z)}{\bar{I}_{\nu_n}(z)}  \notag \\
	&&\times \left[ e^{-\nu_n\pi i}f(iz)+e^{\nu_n\pi i}f(-iz)\right] ,  \label{sumform1AP}
\end{eqnarray}%
where $I_{\nu}(z)$ and $K_{\nu}(z)$ are the modified Bessel functions \cite{Abra}.

For our case, we have
\begin{equation}
	\label{f_function}
f(z)=\frac{zJ_{\nu_n}(z r/a)J_{\nu_n}(zr'/a)} {\sqrt{(z/a)^{2\xi}+\mu^{2\xi}}} e^{-iE_\sigma (t-t')} \ .
\end{equation}

Now, substituting \eqref{f_function} into \eqref{sumform1AP}, we see that the Wightman function can be expressed as the sum of two contributions: a boundary-free contribution and a boundary-induced one, as shown below,
\begin{eqnarray}
	\label{W_csb}
W(x,x')=W_{q}(x,x')+W_b(x,x') \ .
\end{eqnarray}

The boundary-free term, $W_{q}(x,x')$, is given by the first integral and reads
\begin{eqnarray}
	\label{W_function_q}
W_{q}(x,x')&=&\frac{q}{4\pi l^{\xi-1}}\sum_{n=-\infty}^{+\infty}e^{iqn(\varphi-\varphi')}\int_0^\infty d\lambda \lambda \frac{J_{q|n+\alpha|}(\lambda r) J_{q|n+\alpha|}(\lambda r')} {\sqrt{\lambda^{2\xi}+\mu^{2\xi}}}\nonumber\\
&\times& e^{-il^{\xi-1}\sqrt{\lambda^{2\xi}+\mu^{2\xi}}(t-t')} \  .
\end{eqnarray}

The boundary-induced contribution, $W_b(x,x')$, is given by the second integral in \eqref{sumform1AP}. Obtaining this contribution requires more delicate calculations. After using relations involving Bessel functions of imaginary argument \cite{Abra}, we obtain:
\begin{eqnarray}
	\label{Boundary_W_function}
	W_b(x,x')&=&-\frac{iq}{4\pi^2l^{\xi-1}}\sum_{n=-\infty}^\infty e^{iqn(\varphi-\varphi')}\int_0^\infty d\lambda \lambda \frac{{\bar{K}}_{q|n+\alpha|}(a\lambda)} {{\bar{I}}_{q|n+\alpha|}(a\lambda)}I_{q|n+\alpha|}(\lambda r)I_{q|n+\alpha|}(\lambda r')\nonumber\\
	&\times&\left[\frac{e^{-il^{\xi-1}\sqrt{(e^{i\pi/2}\lambda)^{2\xi}+\mu^{2\xi}}(t-t')}}{\sqrt{(e^{i\pi/2}\lambda)^{2\xi}+\mu^{2\xi}}}-\frac{e^{-il^{\xi-1}\sqrt{(e^{-i\pi/2}\lambda)^{2\xi}+\mu^{2\xi}}(t-t')}}{\sqrt{(e^{-i\pi/2}\lambda)^{2\xi}+\mu^{2\xi}}}\right] .
\end{eqnarray}

Using the identity
\begin{eqnarray}
	\label{Identity}
\sqrt{(e^{\pm i\pi/2}\lambda)^{2\xi}+\mu^{2\xi}}=
	\left\{
	\begin{array}{ll}
\sqrt{\mu^{2\xi}+(-1)^\xi\lambda^{2\xi}} \  \ {\rm for} \  \mu>\lambda \  , & \\
e^{\pm i\pi\xi/2}\sqrt{\lambda^{2\xi}+(-1)^\xi\mu^{2\xi}}  \ {\rm for} \  \mu<\lambda \  , & \\
	\end{array}
	\right.
\end{eqnarray}
we divide the integral over $\lambda$ into two intervals: $[0,\mu]$ and $[\mu,\infty)$. The contributions from the first interval cancel each other, leaving only the integral over the second one. Thus, after some minor steps, we obtain the following expression for the boundary-induced Wightman function:
\begin{eqnarray}
	\label{W_function_in}
	W_b(x,x')&=&-\frac{q\sin(\pi \xi/2)}{2\pi^2 l^{\xi-1}}\sum_{n=-\infty}^{+\infty}e^{iqn(\varphi-\varphi')} \int_{\mu}^\infty d\lambda \lambda \frac{{\bar{K}}_{q|n+\alpha|}(a\lambda)} {{\bar{I}}_{q|n+\alpha|}(a\lambda)} \nonumber\\
	&\times& I_{q|n+\alpha|}(\lambda r)I_{q|n+\alpha|}(\lambda r')
\frac{\cosh\!\left[l^{\xi-1}\sqrt{\lambda^{2\xi}-\mu^{2\xi}}\,(t-t')\right]}
{\sqrt{\lambda^{2\xi}-\mu^{2\xi}}} \ .
\end{eqnarray}

Due to the asymptotic behavior of the modified Bessel functions \cite{Abra}, we can see that, in the limit $a\rightarrow\infty$ with fixed $r$ and $r'$, $W_b(x,x')$ vanishes; consequently, the Wightman function is given solely by $W_q(x,x')$.

Also, an important consequence of Eq.~\eqref{W_function_in} is that the boundary-induced contribution vanishes identically for even values of the critical exponent $\xi$, due to the factor $\sin(\pi\xi/2)$. Therefore, only odd values of $\xi$ lead to nonvanishing boundary effects in the Wightman function.

\subsubsection{Outside region}
\label{ext_reg}
In the region outside the circle, the radial solution of the wave-function is expressed as a combination of Bessel and Neumann functions, as shown below:
\begin{eqnarray}
	\label{W-function}
	W_{\nu_n}(\lambda r)=C_1J_{q|n+\alpha|}(\lambda r)+C_2Y_{q|n+\alpha|}(\lambda r) \  .
\end{eqnarray}

Due to the imposition of the Robin boundary condition, Eq.~\eqref{Dirbc}, on the above radial solution, we obtain the relation
\begin{eqnarray}
C_2/C_1=-\bar{J}_{q|n+\alpha|}(\lambda a)/\bar{Y}_{q|n+\alpha|}(\lambda a) \  .	
\end{eqnarray}

So, we can write the solution as
\begin{eqnarray}
	\label{Out_sol}
\phi_\sigma(x)=C_\sigma g_{q|n+\alpha|}(\lambda r,\lambda a)e^{-i(E t-qn\phi )} \ .
\end{eqnarray}
In the above expression, we have introduced the definition
\begin{equation}
	\label{g-function}
	g_{\nu}(u,v)=J_\nu(v)\bar{Y}_{\nu}(u)-\bar{J}_{\nu}(u)Y_{\nu}(v) \ .
\end{equation}
In the region outside the circle, the parameter $\lambda$ assumes continuous values; so the normalization condition for this case reads,
\begin{eqnarray}
	\label{Norm2}
	\int_a^\infty d^2x\sqrt{|g|}\phi^*_{\sigma'}(x)\phi_\sigma(x)=\frac{\delta(\lambda-\lambda')\delta_{n,n'}}{2E_\sigma} \  .
\end{eqnarray}
To obtain the normalization coefficient, we observe that the integral diverges in the limit
$\lambda=\lambda^{\prime}$, so the main contribution to the integral comes from large values of $r$, and in this case we can replace the Bessel functions by their asymptotic expressions for large arguments. This procedure enables us to obtain the integral, providing the following result for the normalization constant:
\begin{equation}
	|C_\sigma|^2=\frac{q}{4\pi E_\sigma}\frac \lambda{\bar{J}_{q|n+\alpha|}^2(\lambda a)+\bar{Y}_{q|n+\alpha|}^2(\lambda a)} \ .
	\label{beta2}
\end{equation}

Substituting \eqref{Out_sol} into the expression \eqref{W.function}, the positive-frequency Wightman function can be expressed as
\begin{eqnarray}
	\label{W_outside}
W(x,x')&=&\frac{q}{4\pi l^{\xi-1}}\sum_{n=-\infty}^\infty e^{iqn(\varphi-\varphi')}\int_0^\infty d\lambda \lambda \frac1{\bar{J}_{q|n+\alpha|}^2(\lambda a)+\bar{Y}_{q|n+\alpha|}^2(\lambda a)} \nonumber\\
&\times&\frac1{\sqrt{\lambda^{2\xi}+\mu^{2\xi}}}	g_{q|n+\alpha|}(\lambda r,\lambda a)	g_{q|n+\alpha|}(\lambda r',\lambda a)e^{-iE_\sigma(t-t')}  \ .
\end{eqnarray}
At this point, we will use the identity below:
\begin{equation}
\frac{g_{\nu_n}(\gamma r,\gamma a)g_{\nu_n}(\gamma r^{\prime },\gamma a)}{\bar{J}_{\nu_n}^{2}(\gamma a)+\bar{Y}_{\nu_n}^{2}(\gamma a)}=J_{\nu_n}(\gamma r)J_{\nu_n}(\gamma r^{\prime })-\frac{1}{2}\sum_{l=1}^{2}\frac{\bar{J} _{\nu_n}(\gamma a)}{{\bar H}_{\nu_n}^{(l)}(\gamma a)}H_{\nu_n}^{(l)}(\gamma r)H_{\nu_n}^{(l)}(\gamma r^{\prime }),  \label{relext}
\end{equation}%
where $H_{\nu}^{(l)}(z)$, for $l=1,2$, are the Hankel functions \cite{Abra}. In this way, we can express the Wightman function as the sum of a boundary-free contribution in the form \eqref{W_csb}, with its boundary-induced part being given by
\begin{eqnarray}
W_b(x,x')&=&-\frac{q}{8\pi l^{\xi-1}}\sum_{n=-\infty}^{+\infty} e^{iqn(\varphi-\varphi')} \int_0^\infty d\lambda \lambda \sum_{l=1}^{2}\frac{\bar{J} _{q|n+\alpha|}(\lambda a)}{{\bar H}_{q|n+\alpha|}^{(l)}(\lambda a)}\nonumber\\
 &\times&H_{q|n+\alpha|}^{(l)}(\lambda r)H_{q|n+\alpha|}^{(l)}(\lambda r^{\prime }) \frac{e^{-il^{\xi-1}\sqrt{\lambda^{2\xi}+\mu^{2\xi}}(t-t')}}{\sqrt{\lambda^{2\xi}+\mu^{2\xi}}} \ .
\end{eqnarray}

In order to proceed with our development, we rotate the integration contour in the complex $\lambda$ plane as follows: by the angle $\pi /2$ for $l=1$ and by the angle $-\pi /2$ for $l=2$. In what follows, we use the relations involving Bessel functions with imaginary arguments \cite{Abra}. Denoting the integral over $\lambda$ by ${\cal I}$, we have:
\begin{eqnarray}
	{\cal{I}}&=&\frac{2i}{\pi}\int_0^\infty d\lambda \lambda \frac{\bar{I}_{q|n+\alpha|}(\lambda a)} {\bar{K}_{q|n+\alpha|}(\lambda a)}K_{q|n+\alpha|}(\lambda r) K_{q|n+\alpha|}(\lambda r')\nonumber\\
	&\times&\left[\frac{e^{-il^{\xi-1}\sqrt{(e^{i\pi/2}\lambda)^{2\xi}+\mu^{2\xi}}(t-t')}}{\sqrt{(e^{i\pi/2}\lambda)^{2\xi}+\mu^{2\xi}}}-\frac{e^{-il^{\xi-1}\sqrt{(e^{-i\pi/2}\lambda)^{2\xi}+\mu^{2\xi}}(t-t')}}{\sqrt{(e^{-i\pi/2}\lambda)^{2\xi}+\mu^{2\xi}}}\right]  \  .
\end{eqnarray}
Now, using the identity \eqref{Identity}, we divide the integrals over $\lambda$ into two segments: $[0, \ \mu]$ and  $[\mu, \ \infty )$. As in the previous subsection, the integrals over the first segment cancel each other, leaving only the integral over the second one. So, our final result for the Wightman function is:
 \begin{eqnarray}
\label{W_function_out}
	W_b(x,x')&=&-\frac{q\sin(\pi \xi/2)}{2\pi^2 l^{\xi-1}}\sum_{n=-\infty}^{+\infty}e^{iqn(\varphi-\varphi')} \int_{\mu}^\infty d\lambda \lambda \frac{{\bar{I}}_{q|n+\alpha|}(a\lambda)} {{\bar{K}}_{q|n+\alpha|}(a\lambda)} \nonumber\\
&\times& K_{q|n+\alpha|}(\lambda r)K_{q|n+\alpha|}(\lambda r')\frac{\cosh[l^{\xi-1}\sqrt{\lambda^{2\xi}-\mu^{2\xi}})(t-t')]} {\sqrt{\lambda^{2\xi}-\mu^{2\xi}}} \  .
 \end{eqnarray}
As we can observe, the above boundary-induced Wightman function is similar to the corresponding one for the region inside, Eq.~\eqref{W_function_in}. The main difference resides in the exchange of the function $I_\nu(z)$ by the function $K_\nu(z)$ and vice versa. Moreover, due to the asymptotic behavior of these Bessel functions, in the limit $a\rightarrow 0$ for fixed $r,r^{\prime }$, \eqref{W_function_out} goes to zero. The same conclusion applies to the exterior region.
For even values of $\xi$, the factor $\sin(\pi\xi/2)$ vanishes and,
consequently, the boundary-induced contribution to the Wightman function
is identically zero.

\section{Boundary-induced current}
\label{sec3}
The expression for the current density operator associated with a bosonic field is given by
\begin{eqnarray}
	\hat{j}_{\mu }(x)&=&ie\left[{\hat\phi} ^{\dagger}(x)D_{\mu }\hat{\phi} (x)-
	(D_{\mu }\hat{\phi)}^{\dagger}\hat{\phi}(x)\right] \nonumber\\
	&=&ie\left[\hat{\phi}^{\dagger}(x)\partial_{\mu }\hat{\phi} (x)-{\hat\phi}(x)
	(\partial_{\mu }\hat{\phi(x))}^{\dagger}\right]-2e^2A_\mu(x)|\hat{\phi}(x)|^2 \   .
	\label{J.mu}
\end{eqnarray}

The main objective of this section is to obtain the VEV of the bosonic current in the regions inside and outside the circle. This quantity can be evaluated by using the positive-frequency Wightman function:
\begin{equation}
	\label{Induced_current}
	\langle {\hat{j}}_{\mu}(x) \rangle=ie\lim_{x'\rightarrow x}
	\left\{(\partial_{\mu}-\partial_{\mu '})W(x,x')+2ieA_\mu W(x,x')\right\} \ .
\end{equation}

For the system under consideration, the only non-vanishing component of the current density is the azimuthal one. So our focus will be on obtaining this component.

Taking the expression, $A_\varphi=q\frac{\alpha}{e}$, for the three-vector potential, the azimuthal current density is expressed as
\begin{eqnarray}
\label{Ind-Curr}
\langle {\hat{j}}_{\varphi}(x) \rangle=2ie\lim_{x'\rightarrow x}	\left\{\partial_{\varphi}W(x,x')+iq\alpha W(x,x')\right\} \ .
\end{eqnarray}

By using the general expressions for the Wightman functions for the regions inside and outside the circle, Eq. \eqref{W_csb}, the bosonic induced current is given as the sum of a boundary-free current plus the boundary-induced one:
\begin{eqnarray}
	\langle {\hat{j}}_{\varphi}(x) \rangle=\langle {\hat{j}}_{\varphi}(x) \rangle_q +\langle {\hat{j}}_{\varphi}(x) \rangle_b \ .
\end{eqnarray}

Because the boundary-free induced current presents the same structure for both regions, here we evaluate this component and leave the next subsections to evaluate the boundary-induced components in the regions inside and outside the circle. So, substituting \eqref{W_function_q} into \eqref{Ind-Curr}, after some direct intermediate steps, we obtain:
\begin{eqnarray}
	\label{Ind_curr_1}
	\langle {\hat{j}}_{\varphi}(x) \rangle_q=-\frac{eq^2}{2\pi r} \left(\frac r l\right)^{\xi-1} \sum_{n=-\infty}^\infty(n+\alpha)\int_0^\infty dz z \frac{J^2_{q|n+\alpha|}(z)}{\sqrt{z^{2\xi}+(r\mu)^{2\xi}} }\  ,
\end{eqnarray} 
where we have changed the integration variable to $\lambda=z/r$.\footnote{Note that  $\mu r$ in \eqref{Ind_curr_1} can be expressed as $\mu r=(mr)^{1/\xi}(r/l)^{1-1/\xi}=(mr)/(ml)^{1-1/\xi}$. }

We can see that this current is an odd function of the magnetic flux. A more convenient expression for \eqref{Ind_curr_1} can be obtained by expressing the parameter $\alpha$ defined in \eqref{alpha_para} in the form
\begin{equation}
	\alpha=n_0 +\alpha_0 \ \ {\rm with  \ |\alpha_0| \ < 1/2},
	\label{alphazero}
\end{equation}
where $n_0$ is an integer number. Redefining the quantum number $n$ as $n\to n-n_0$, the above expression will depend only on $\alpha_0$, that is, the ratio between the total magnetic flux, $\Phi$, and the quantum one, $\frac{2\pi}{e}$.

\begin{figure}[!htb]
	\begin{center}
		\includegraphics[scale=0.40]{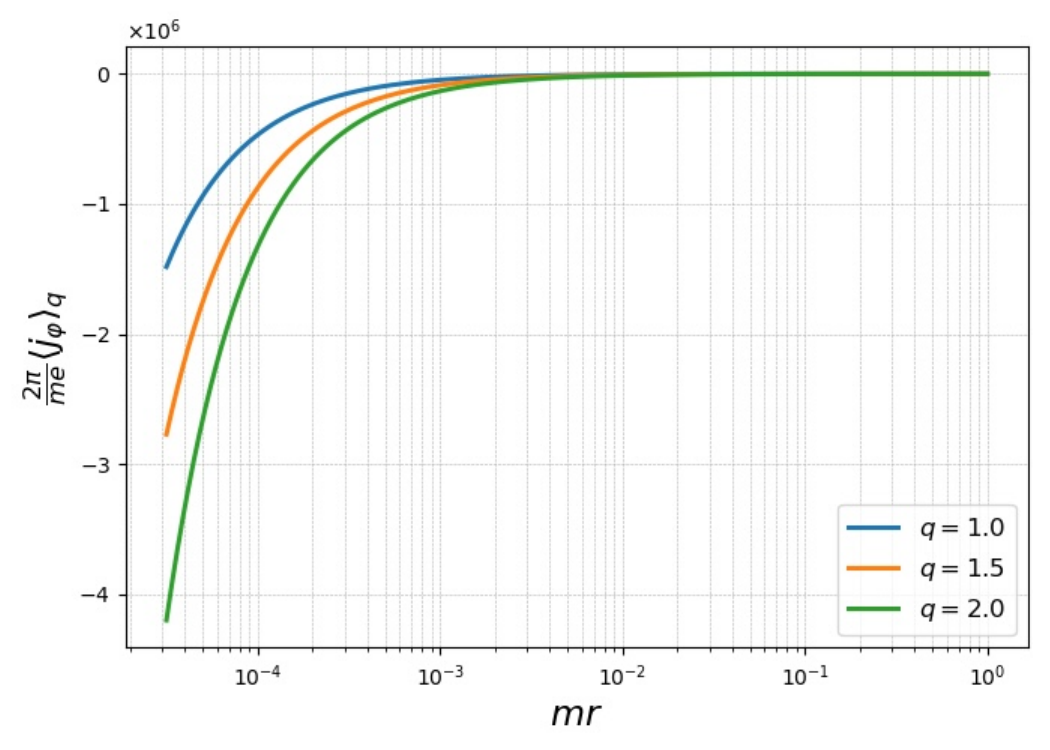}
		\vfill
		\includegraphics[scale=0.40]{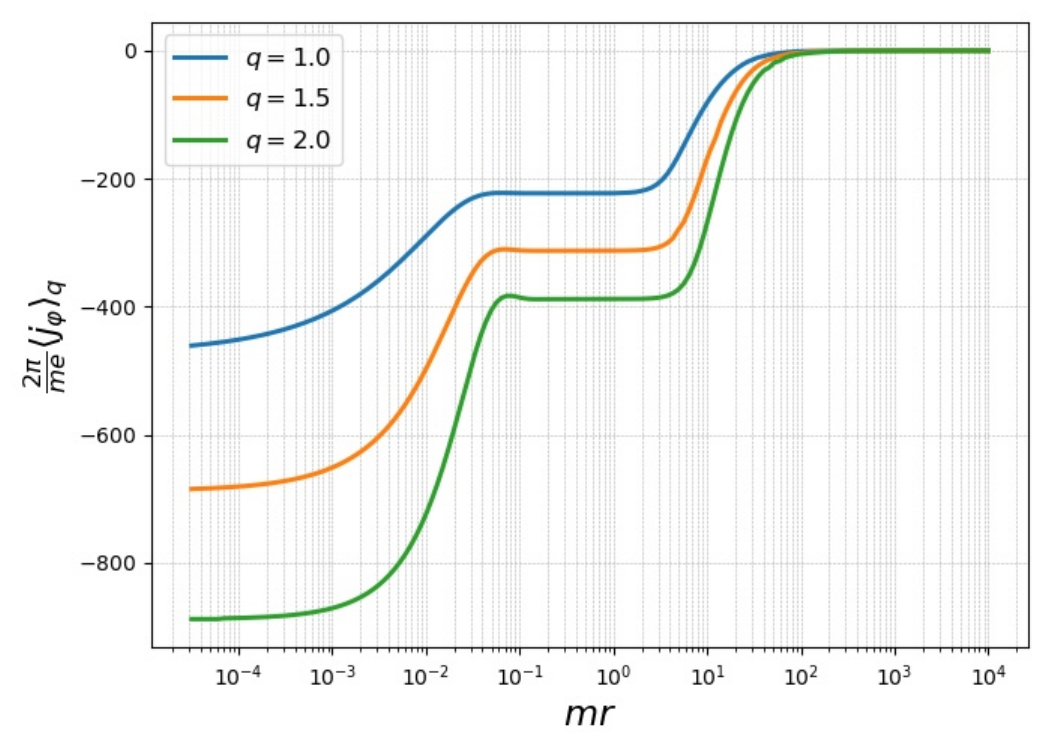}
		\vfill
		\includegraphics[scale=0.40]{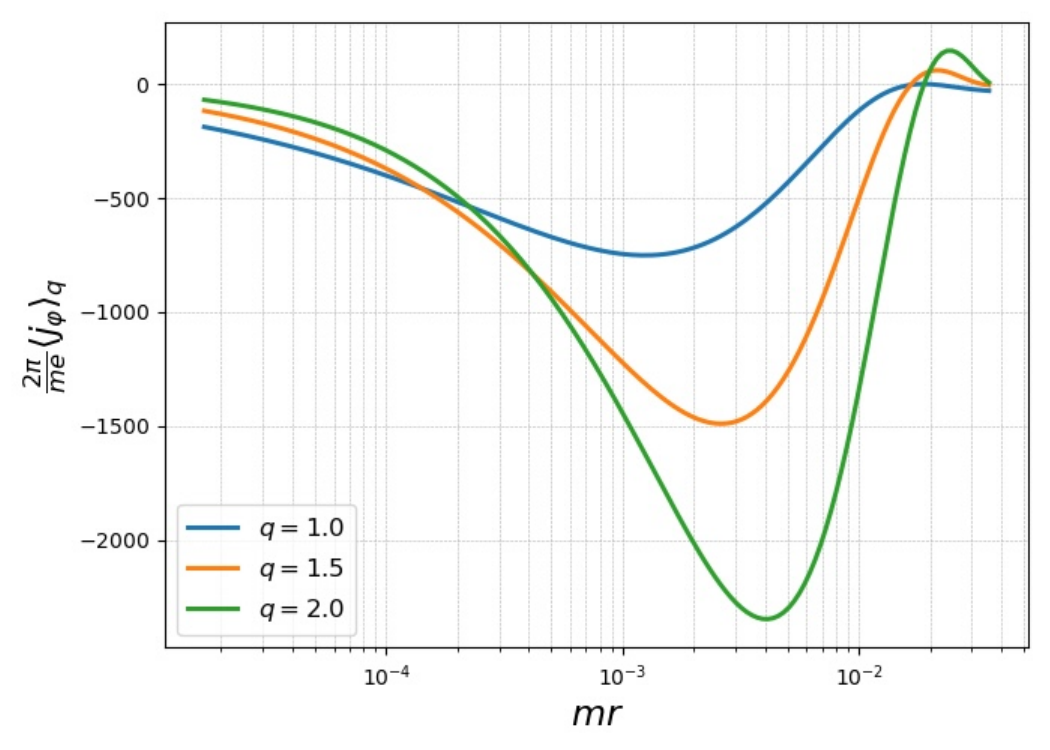}
		\caption{These plots exhibit the behavior of the boundary-free induced current in units of $me$ as a function of $mr$, considering different values of the critical exponent: $\xi=1$ (upper), $\xi=2$ (middle), and $\xi=3$ (bottom). For each plot, we assumed $q=1.0, \ 1.5, \ 2.0$. Moreover, we adopted $\alpha_0=1/4$ and $ml=10^{-3}$. Note that the values on the horizontal axis are displayed on a logarithmic scale.}
		\label{fig1}
	\end{center}
\end{figure}

In \cite{Braganca2015} a closed expression for the bosonic current can be found for $\xi=1$; however, for $\xi\geq 2$ the integral over the variable $z$ presents a long and unenlightening result. In order to provide a better understanding of the induced current given in \eqref{Ind_curr_1}, we exhibit in Fig. \ref{fig1} the behavior of $\frac{\langle {\hat{j}}_{\varphi}(x) \rangle_q}{me}$ as a function of $mr$, for different values of $q$ (the values are specified in the plots), considering $\xi=1$, in the upper plot, $\xi=2$, in the middle plot, and $\xi=3$ in the bottom plot. In these plots we consider $\alpha_0=1/4$ and $ml=10^{-3}$.

By these plots we can see that the intensity of the induced current increases with the parameter $q$, for each value of $\xi$. This is a usual behavior. However, we would like to highlight the peculiar behavior of the induced current near the magnetic flux. For $\xi=1$ its behavior is the usual one. It is divergent, going to zero for points far from the core; however, for $\xi=2$ the current near the core becomes finite, and mainly, for $\xi>2$, the current goes to zero at the center of the magnetic flux. These behaviors are new. They are consequences of the pre-factor $(r/l)^{\xi-1}$ in front of \eqref{Ind_curr_1}. Also, we can see that the intensity of the current decreases for large values of $\xi$.

\subsection{Boundary induced current in the region inside}
\label{Boudary_in}
The boundary-induced current in the region inside the circle can be obtained by substituting \eqref{W_function_in} into \eqref{Ind-Curr}. After some intermediate steps, we get,
\begin{eqnarray}
	\label{Curr_in}
\langle {\hat{j}}_{\varphi}(x) \rangle_{b}^{(in)}&=&\frac{eq^2}{a\pi^2}\left(\frac al\right)^{\xi-1}\sin(\pi\xi/2) \sum_{n=-\infty}^{+\infty}(n+\alpha_0)\nonumber\\
&\times& \int_{a\mu}^\infty dz z \frac{{\bar{K}}_{q|n+\alpha_0|}(z)}{{\bar{I}}_{q|n+\alpha_0|}(z)}\frac{I^2_{q|n+\alpha_0|}(z(r/a))}{\sqrt{z^{2\xi}-(a\mu)^{2\xi}}} \ ,
\end{eqnarray}
where we have also changed the integral variable $\lambda$ by $z/a$ in \eqref{W_function_in}. We therefore conclude that the boundary-induced current exists only for odd values of $\xi$. For even values of the critical exponent, the current is completely determined by the boundary-free contribution.
\footnote{In \eqref{Curr_in}, we use $a\mu=(am)^{1/\xi}(a/l)^{1-1/\xi}$.}

In a previous publication, \cite{deMello:2025ajn}, we have analyzed the behavior of the boundary-induced azimuthal current in $(1+D)-$dimensional cosmic string spacetime in the presence of a cylindrical boundary co-axial to the string, considering a Lorentz-symmetry-preserved scenario. The obtained expression for the boundary-induced current coincides with \eqref{Curr_in} for $\xi=1$, taking $D=2$ in the former result. Moreover, in \cite{deMello:2025ajn} it was shown that $\langle {\hat{j}}_{\varphi}(x) \rangle_{b}^{(in)}$ diverges near the boundary as $\frac1{(a-r)^{D-1}}$. In order to analyze this behavior we had to consider large values of $n$. In this limit
the order of the modified Bessel function can be approximated by $n+\alpha_0\approx n$. Due to the presence of $\alpha_0\neq 0$, there remains a non-vanishing contribution in the summation over $n$, proportional to this parameter. We have also introduced a new integration variable $z\rightarrow nqz$, replaced the modified Bessel functions by their uniform asymptotic expansions for large values of the order \cite{Abra}, and expanded the result over $a-r$, up to the leading order. \footnote{See also \cite{Mello} for more details.} Here, to analyze the behavior of \eqref{Curr_in} near the boundary, we also apply a similar procedure; however, for this present analysis, we cannot take the massless limit, because the above expression diverges at the origin for $\xi\geq2$. Again, defining a new variable $z\rightarrow \nu z$, being $\nu=qn$, we get:
\begin{eqnarray}
	\label{Curr_in_1}
	\langle {\hat{j}}_{\varphi}(x) \rangle_{b}^{(in)}&\approx& \frac{2\alpha_0 e q}{a\pi}\left(\frac al\right)^{\xi-1}\sin(\pi\xi/2)\sum_{n=1}^\infty \nu^{2-\xi}\int_{a\mu/\nu}^\infty dz z \frac1{\sqrt{z^{2\xi}-(a\mu/\nu)^{2\xi}}}\nonumber\\
	&\times&\frac{{\bar{K}}_{\nu}(\nu z)}{{\bar{I}}_{\nu}(\nu z)}I^2_{\nu}(\nu z(r/a)) \  .
\end{eqnarray}
Using the uniform asymptotic expansion for large orders of the Bessel function \cite{Abra}, we get:
\begin{eqnarray}
\frac{{\bar{K}}_{\nu}(\nu z)}{{\bar{I}}_{\nu}(\nu z)}I^2_{\nu}(\nu z(r/a))\approx \frac{e^{-2\nu(\eta(z)-\eta(z(r/a)))} }{2\nu}\frac{(\delta_{B,0}-1)}{\sqrt{1+z^2(r/a)^2}}  \  ,
\end{eqnarray}
with
\begin{eqnarray}
	\eta(z)=\sqrt{1+z^2}+\ln\left(\frac{z}{1+\sqrt{1+z^2}}\right) \  .
\end{eqnarray}
However, for $r$ close to $a$, we get,
\begin{eqnarray}
\eta(z)-\eta(z(r/a))	\approx \sqrt{1+z^2}(1-r/a) \  .
\end{eqnarray}
Now changing $z=x(a\mu/\nu)$ in \eqref{Curr_in_1}, we have,
\begin{eqnarray}
	\label{Approx}
	\langle {\hat{j}}_{\varphi}(x) \rangle_{b}^{(in)}&\approx&\frac{\alpha_0 e qm(ml)^{2/\xi-2}} {\pi}\sin(\pi\xi/2)(\delta_{B,0}-1)\int_{1}^\infty dx\frac{x} {\sqrt{x^{2\xi}-1}}\nonumber\\
	&\times&\sum_{\nu=1}^\infty \frac{e^{-2\sqrt{\nu^2+(a\mu x)^2}(1-r/a)}}{\sqrt{\nu^2+(a\mu x)^2}} \  .
\end{eqnarray}
After approximately evaluating the sum over $\nu$ and performing the integral in the variable $x$, we obtain,
\begin{eqnarray}
	\label{Approx-1}
		\langle {\hat{j}}_{\varphi}(x) \rangle_{b}^{(in)}&\approx& -\frac{\alpha_0 qm }{2\xi\sqrt{\pi}}(ml)^{2/\xi-2}\sin(\pi\xi/2) (\delta_{B,0}-1)\ln(2a\mu(1-r/a))\frac{\Gamma \left( {\frac {\xi-2}{2\xi}
			} \right) }{\Gamma \left(\frac {\xi-1}{\xi}\right) 	}  \  ,
\end{eqnarray}
valid only for $\xi>2$ and $m\neq 0$. The logarithmic divergence exhibited in Eq.~(50) should be interpreted as
a surface divergence associated with the idealized implementation of the
Robin boundary condition. Similar divergences are well known in Casimir
problems involving perfectly reflecting boundaries and do not correspond
to ultraviolet divergences of the bulk vacuum current.

\begin{figure}[!htb]
	\begin{center}
		\includegraphics[scale=0.30]{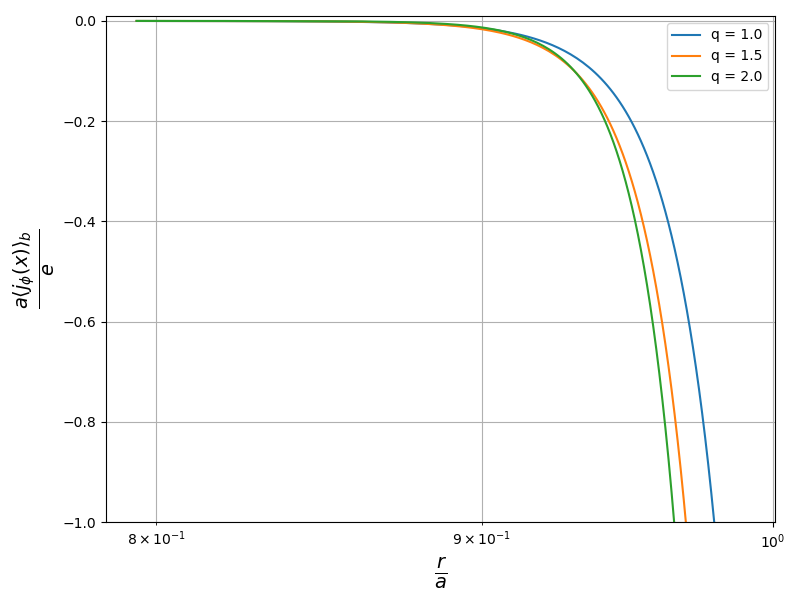}
		\quad
		\includegraphics[scale=0.30]{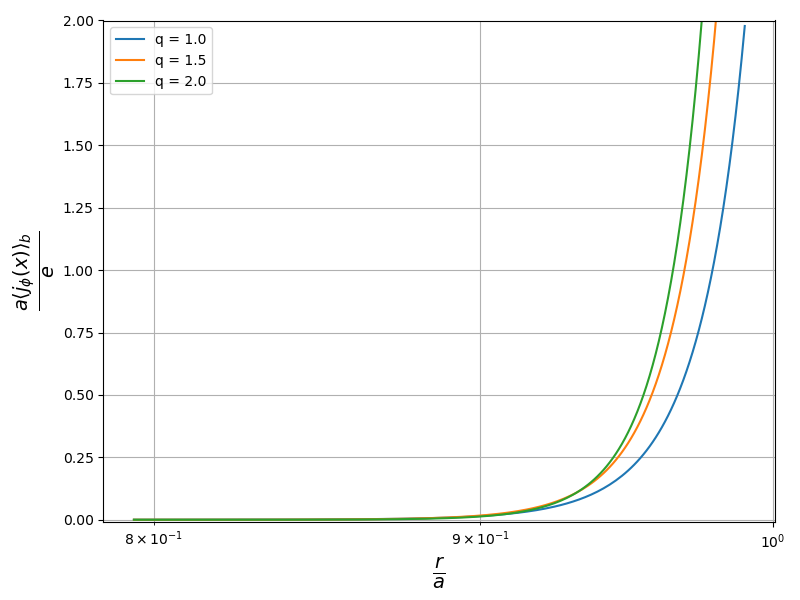}
		\vfill
		\includegraphics[scale=0.30]{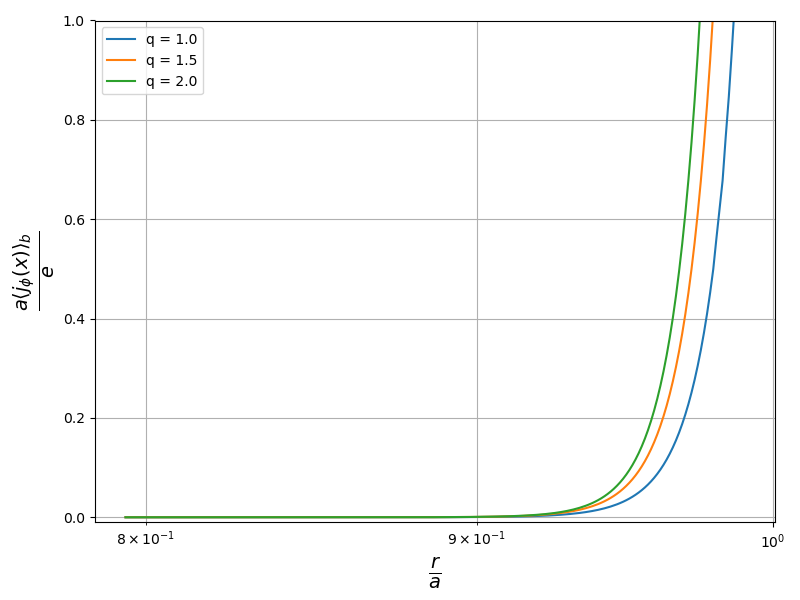}
		\quad
		\includegraphics[scale=0.30]{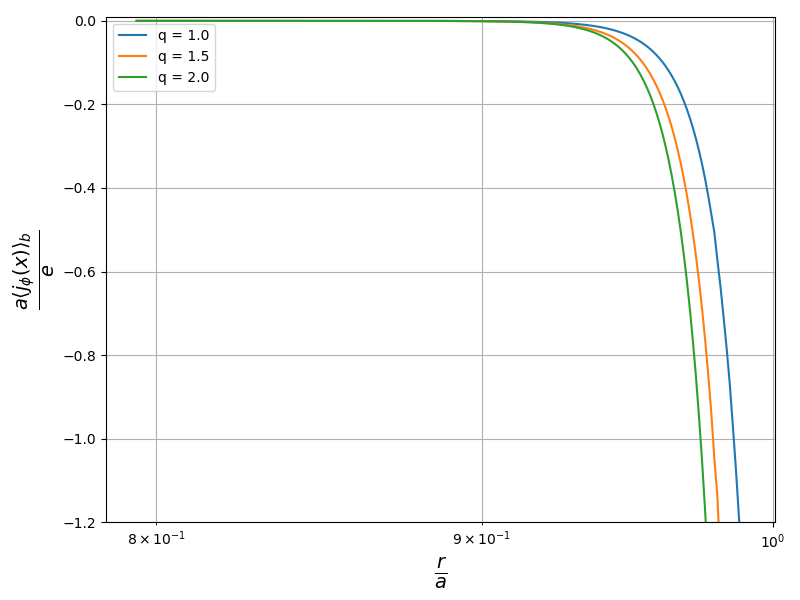}
		\caption{These plots exhibit the boundary-induced behavior of $\frac{a\langle {\hat{j}}_{\varphi}(x) \rangle_{b}^{(in)}}{e}$ as a function of $(r/a)$ in the region inside the circle, considering the Dirichlet boundary condition for the left plots and the Neumann boundary condition for the right plots. The upper plots are for $\xi=3$ and the lower ones for $\xi=5$. Different values of $q$ (the numbers near the curves) are adopted; moreover, we assume $\alpha_0=1/4$, $(a/l)=10^2$ and $ml=10^{-2}$. Note that the values on the horizontal axis are on a logarithmic scale.}
		\label{fig2}
	\end{center}
\end{figure}

In Fig. \ref{fig2} the behaviors of the boundary-induced currents in the region inside the circle are exhibited as a function of $r/a$, considering $\xi=3$ for the plots on the top, and $\xi=5$ for the plots on the bottom. For both cases, we take $\alpha_0=1/4$, $(a/l)=10^2$, $ml=10^{-2}$ and different values of $q$. In the left plots the Dirichlet boundary condition is adopted, and in the right ones the Neumann boundary condition.

\subsection{Boundary-induced current outside the shell}
\label{Boudary_out}
The boundary-induced current in the region outside the circle can be obtained from \eqref{Curr_in} by changing $I\rightleftarrows K$. The final result is,
\begin{eqnarray}
	\label{Curr_out}	
	\langle {\hat{j}}_{\varphi}(x) \rangle_{b}^{(out)}&=&\frac{eq^2}{a\pi^2}\left(\frac al\right)^{\xi-1}\sin(\pi\xi/2) \sum_{n=-\infty}^{+\infty}(n+\alpha_0)\nonumber\\
	&\times& \int_{a\mu}^\infty dz z \frac{{\bar{I}}_{q|n+\alpha_0|}(z)}{{\bar{K}}_{q|n+\alpha_0|}(z)}\frac{K^2_{q|n+\alpha_0|}(z(r/a))}{\sqrt{z^{2\xi}-(a\mu)^{2\xi}}} \  .
\end{eqnarray}
Here, we can also verify that for points far away from the circle, $r>a$, this current goes to zero.

As in the region inside the boundary, Eq.~\eqref{Curr_out} also diverges near the cylindrical shell. The leading term can be obtained in the same way as in the previous analysis. The corresponding asymptotic expansion is similar to \eqref{Approx-1}, by changing $\ln(2a\mu(1-r/a))$ to $\ln(2a\mu(r/a-1))$. This divergence is also localized on the boundary and originates from the idealized description of the circular shell. It is important to emphasize that the logarithmic growth of the current near $r=a$, both inside and outside the circle, should not be interpreted as a physical effect. Rather, it signals the breakdown of the idealized boundary description in the immediate vicinity of the shell, which has been modeled as a perfectly reflecting interface.

Now let us investigate the behavior of \eqref{Curr_out} for points very far from the boundary, i.e., $r>>a$. Due to the exponential decay of the Macdonald function, the integral is dominated by the lower limit. Under this condition we can approximate ${\sqrt{z^{2\xi}-(a\mu)^{2\xi}}}\approx \sqrt{2(a\mu)^\xi}\sqrt{z^{\xi}-(a\mu)^{\xi}}$. Defining a new integration variable, $z=a\mu (v+1)$, we can obtain an upper limit for the integral in this new variable $v$. So, after some intermediate steps, we get,
\begin{eqnarray}
\label{Upper_limit}
	\langle {\hat{j}}_{\varphi}(x) \rangle_{b}^{(out)}&\approx&\frac{eq^2m}{4\sqrt{\pi\xi}}\frac{(ml)^{2/\xi-2}}{(\mu r)^{3/2}} \sin(\pi\xi/2)e^{-2\mu r}\sum_{n=-\infty}^{+\infty}(n+\alpha_0)\frac{{\bar{I}}_{q|n+\alpha_0|}(a\mu)}{{\bar{K}}_{q|n+\alpha_0|}(a\mu)} \  ,
\end{eqnarray}
where an exponential decay appears.

\begin{figure}[!htb]
	\begin{center}
		\includegraphics[scale=0.30]{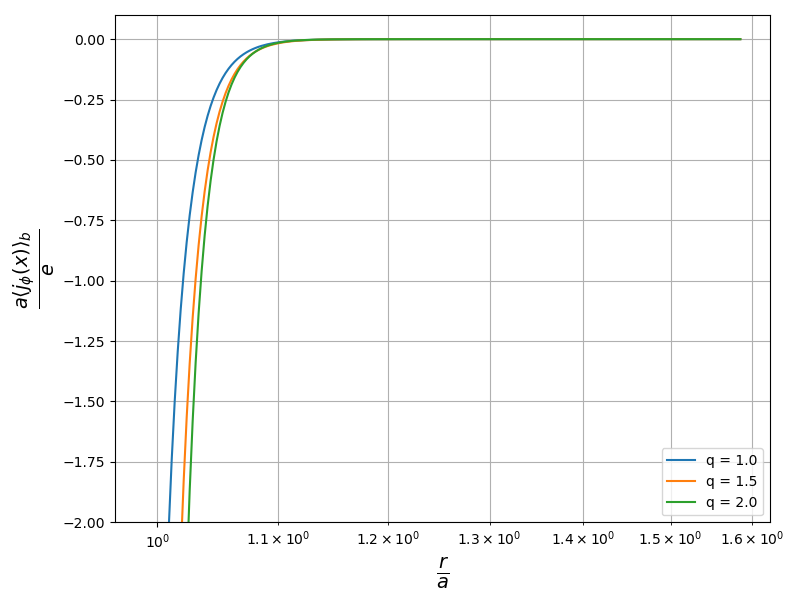}
		\quad
		\includegraphics[scale=0.30]{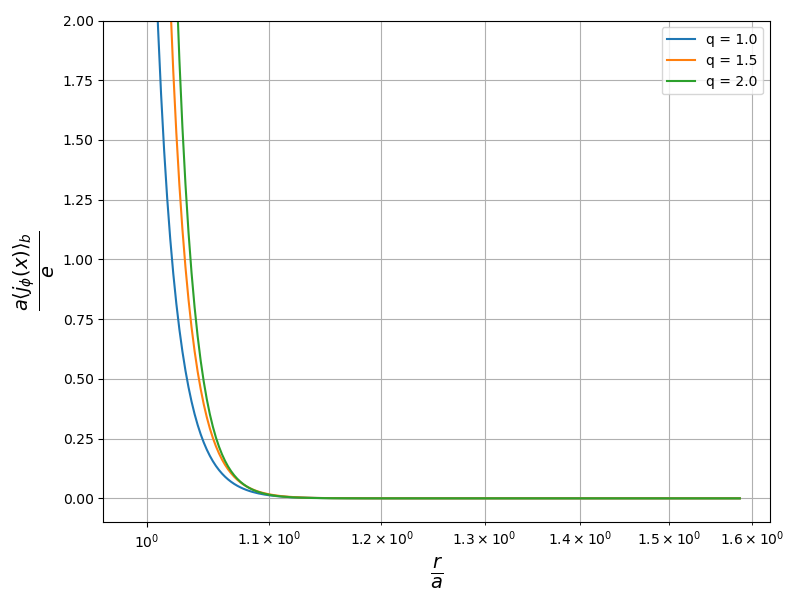}
		\vfill
		\includegraphics[scale=0.30]{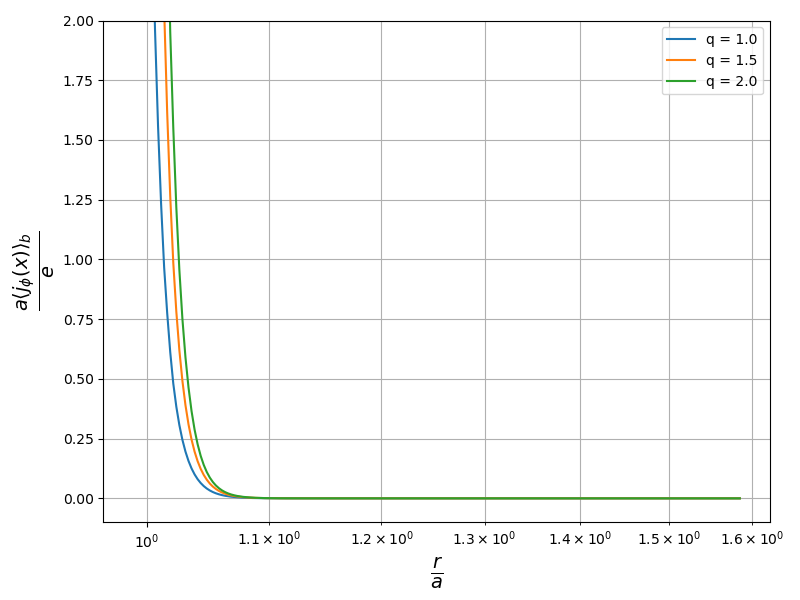}
		\quad
		\includegraphics[scale=0.30]{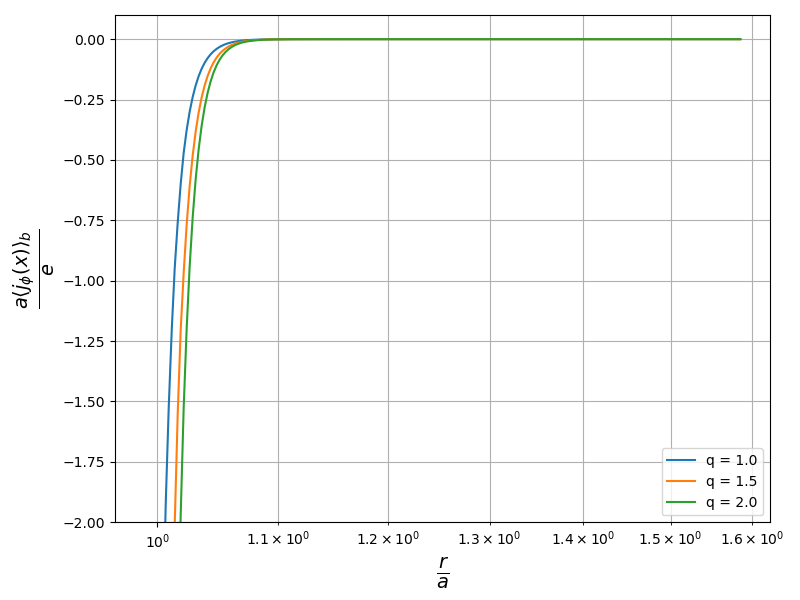}
		\caption{These plots exhibit the boundary-induced behavior of $\frac{a\langle {\hat{j}}_{\varphi}(x) \rangle_{b}^{(in)}}{e}$ as a function of $(r/a)$ in the region outside the circle. We consider the Dirichlet boundary condition for the left plots and the Neumann boundary condition for the right plots. Moreover, the upper plots are for $\xi=3$ and the lower ones for $\xi=5$. Different values of $q$ (the numbers near the curves) are adopted; moreover, we assume $\alpha_0=1/4$, $(a/l)=10^2$ and $ml=10^{-2}$. Note that the values on the horizontal axis are on a logarithmic scale.}
		\label{fig3}
	\end{center}
\end{figure}

In Fig. \ref{fig3} are presented the behaviors of the boundary-induced currents in the region outside the circle as a function of $r/a$. The plots on the top are for $\xi=3$, and the plots on the bottom are for $\xi=5$. For both cases we assume $\alpha_0=1/4$, $(a/l)=10^2$, $ml=10^{-2}$ and different values of $q$. In the left plots we adopted the Dirichlet boundary condition, and in the right ones the Neumann boundary condition.

\section{Conclusions}
\label{conc} 
In this paper, we have investigated the vacuum bosonic current induced by a magnetic flux located at the apex of a (1+2)-dimensional conical spacetime in a Hořava-Lifshitz Lorentz-violating scenario, taking into account the presence of a circular boundary of radius $a$ concentric with the cone. We assumed that the quantum field obeys the Robin boundary condition on the circular boundary. Specifically, we calculated the induced vacuum currents in the regions inside and outside the circle.
To develop this analysis, we first obtained the corresponding positive-frequency Wightman functions, Eqs.~\eqref{W_inside} and \eqref{W_outside}, for the interior and exterior regions, respectively. Both functions were expressed as the sum of two contributions: a boundary-free term and a boundary-induced term. Using the general expression for the induced azimuthal current, Eq.~\eqref{Ind-Curr}, we showed that the corresponding vacuum current can also be decomposed into a boundary-free part, $\langle \hat{j}(x)\rangle_q$, and a boundary-induced part, $\langle \hat{j}(x)\rangle_b$.

 The boundary-free current has the same structure in both regions and is given by Eq.~\eqref{Ind_curr_1}, independently of the value of the critical exponent $\xi$. The boundary-induced currents for the interior and exterior regions are given by Eqs.~\eqref{Curr_in} and \eqref{Curr_out}, respectively. An important feature of these expressions is that the boundary-induced contributions exist only for odd values of $\xi$. This behavior originates from the analytic structure of the integrand in Eq.~\eqref{Boundary_W_function}. After the contour rotation in the complex plane, a non-vanishing contribution arises from the discontinuity across the branch cut $[\mu,\infty)$ only when $\xi$ is odd, leading to the factor $\sin(\pi\xi/2)$ appearing in Eqs.~\eqref{Curr_in} and \eqref{Curr_out}. Consequently, for even values of $\xi$, this factor vanishes identically, and the boundary-induced contributions to both the Wightman function and the vacuum current disappear. Therefore, in the even-$\xi$ sector, the total induced current is entirely determined by the boundary-free contribution.
In the Lorentz-invariant case, $\xi=1$, Casimir-type effects associated with bosonic fields have been investigated in Ref.~\cite{Bordag} using a similar contour integration procedure.

As can be seen, all the results obtained in this paper depend on the critical exponent $\xi$, which is assumed to be an integer greater than unity. As a consequence, even the boundary-free induced current cannot, in general, be expressed in terms of a finite number of elementary functions, in contrast to the Lorentz-invariant case $\xi=1$, for which a closed-form expression is available \cite{Braganca2015}.
To gain further insight into the behavior of the boundary-free current, in Fig.~\ref{fig1} we display the quantity $\langle {\hat{j}}_{\varphi}(x) \rangle_q/(me)$ as a function of $mr$, for different values of the parameter $q$. The upper, middle, and lower panels correspond to $\xi=1$, $\xi=2$, and $\xi=3$, respectively. As shown in the figure, the boundary-free induced current remains finite at the cone apex ($r=0$) for all cases with $\xi \geq 2$.
For the Lorentz-invariant case, $\xi=1$, the current exhibits the same singular behavior at $r=0$ as the azimuthal vacuum current previously obtained in Ref.~\cite{Braganca2015}. This divergence is a consequence of the idealized description adopted for the magnetic flux, which is assumed to be confined to an infinitely thin flux tube located at the apex of the cone.

The analysis of the boundary-induced current in the regions inside and outside the circle was presented in Subsections~\ref{Boudary_in} and~\ref{Boudary_out}, respectively. The corresponding numerical behaviors are displayed in Figs.~\ref{fig2} and \ref{fig3} for the interior and exterior regions, respectively, considering the cases $\xi=3$ and $\xi=5$. In both figures, we have adopted $\alpha_0=1/4$, $(a/l)=10^2$, $ml=10^{-2}$, and different values of $q$. The plots on the left correspond to Dirichlet boundary conditions, whereas those on the right correspond to Neumann boundary conditions. One can clearly observe the change in the sign of the induced current when the boundary condition is changed from Dirichlet to Neumann. In addition, there is a sign reversal between the cases $\xi=3$ and $\xi=5$, which follows directly from the factor $\sin(\pi\xi/2)$ appearing in the corresponding expressions.

We also showed that for points near the conical apex, $r\ll a$, the integral converges exponentially at the upper limit, whereas
for points approaching the boundary from either side,
the idealized model predicts a logarithmic growth of the
currents. Consequently, the value exactly at $r=a$ should not
be interpreted as a physical observable, since the boundary
has been modeled as a perfectly reflecting interface of
zero thickness. Moreover, outside the circle, the boundary-induced current decays exponentially as $e^{-2\mu r}$ for points far away, that is, $r\gg a$.

Finally, we would like to comment on the singular behavior of $\langle\hat{j}(x)\rangle_b$ near the boundary. This divergence should be regarded as a surface divergence arising from the idealized description of the circular boundary as a perfectly reflecting interface. Similar divergences are commonly encountered in Casimir-type problems with perfect boundaries. A possible way to regularize this behavior is to allow the boundary position to fluctuate according to a Gaussian probability distribution, thereby replacing the idealized boundary by an effective smeared interface. In this case, the logarithmic divergence is expected to be smoothed out and the current remains finite in the vicinity of the boundary. This approach has been successfully employed, for instance, in Refs.~\cite{DeLorenci,GuedesMota2025}. 

Furthermore, the logarithmic term displayed in Eq.~\eqref{Approx-1}, and the corresponding one for the exterior region, represent the leading near-boundary contributions, and their coefficients should be regarded as the universal part of the corresponding asymptotic expansions. In the same spirit as surface divergences appearing in standard Casimir configurations with idealized boundaries, these coefficients are independent of the specific regularization procedure adopted to smooth the boundary. As discussed in the previous paragraph, a regulator-dependent finite contribution may be present in a more realistic description of the circular shell; however, such a contribution only modifies the local behavior in the immediate vicinity of the boundary and does not affect the bulk properties of the induced current, nor the parity selection rule governed by the factor $\sin(\pi\xi/2)$.

Hence, we stress that the logarithmic divergence discussed in this work should be understood as a local surface effect associated with the idealized boundary model, and not as a prediction for an experimentally measurable current exactly on the boundary. We leave a more detailed investigation of this issue for future work.

\section*{Acknowledgment}
E.R.B.M. and H.F.S.M. would like to thank the Brazilian National Council for Scientific and Technological Development (CNPq) for partial financial support under Grants No. 304332/2024-0 and No. 308049/2023-3, respectively.

\end{document}